\documentstyle[12pt]{article}

\begin{document}
\vspace*{.3cm}
\begin{flushright}
\large{CINVESTAV-FIS-16-95}
\end{flushright}
\vspace{1.0cm}
\begin{center}
\LARGE{\bf T-violation in $K_{\mu3}$ decay in a general two-Higgs doublet
model.}
\end{center}
\vspace{.8cm}
\begin{center}\Large
J. J. Godina Nava \\
\vspace*{.4cm}
{\normalsize{\it Departamento de F\'\i sica, CINVESTAV del IPN,\\
\it Apdo. Postal 14-740, 07000 M\'exico, D.F. M\'EXICO}}
\vspace{.3cm}
\end{center}
\thispagestyle{empty}
\vspace{.3cm}
\centerline{ \bf Abstract}
\vspace{.5cm}

 We calculate the transverse muon polarization in the $K^+_{\mu3}$
process arising from the Yukawa couplings of charged Higgs boson in a
general two-Higgs doublet model where spontaneous violation of CP is
present.

\vspace{1.7cm}
\vspace{5cm}

\newpage
\setcounter{page}{1}
\vspace{2cm}

It has been suggested [1] that the muon polarization transverse
($P^{\perp}_{\mu}$) to the decay plane in the process $K^+\to
\pi^0\mu^+\nu_{\mu}$
($K^+_{\mu3}$) is a good prospect for a measurement [2] of T-violation
(CP is violated as well if T is violated).
The reason for this is that the Standard Model (SM) contribution to this
observable is very suppressed.
Since $P^{\perp}_{\mu}\sim {\cal O}(10^{-6})$ [3] arises from
electromagnetic radiative corrections in the SM, a measurement of a
larger value would signal a violation of T of non-standard origin.

A preliminary measurement of the transverse polarization of the muon
indicates  $P^{\perp}_{\mu}=(-1.85\pm3.6)\times 10^{-3}$ [4], which is
consistent with zero. Thus, a confirmation of $P^{\perp}_{\mu}\sim {\cal
O}(10^{-3})$ would be an interesting tool to test physics beyond the
standard model.
For instance, planed experiments at KEK [2] would be sensitive to
measurements of $P^{\perp}_{\mu}$ at the $5\times 10^{-4}$ level.

The calculation of this observable has been done in models containing
leptoquarks [5], three doublets of Higgses [6], and tensor interactions [7].
In this brief report, which should be regarded as a complement to our
previous work [8], we consider the calculation of $P^{\perp}_{\mu}$ in
the context of a general two-Higgs doublet model where spontaneous
violation of CP is allowed.
In Ref. [8] we have considered only CP-conserving quantities in order to
constrain the absolute values of the Yukawa couplings of the charged
Higgs boson in the model.
In the present work  we consider the calculation of $P^{\perp}_{\mu}$
which would provide a bound on the phase $\delta$ that violates CP (see
below).
This would complete the set of constraints on the additional parameters
of the model.

The part of the model relevant for our calculation is the Lagrangian for
the Yukawa interactions of the charged Higgs boson, namely (see Ref. [8]
for details):
\vspace{.3cm}
\begin{eqnarray}
L_{f_i\bar f_jH^{\pm}}&={\displaystyle \frac {g}{\sqrt{2}m_W}}H^+\bar {U}
[\cot\beta V^+_L M_d R  + \tan\beta V_{L}M_u L \nonumber \\[1em]
&+ \xi e^{-i\delta_1} M_1\Gamma L + \xi e^{-i\delta_2}M_2\Gamma'R ]D
\nonumber \\
& + H^+\overline {\ell_i}[\cot\beta M_{\ell}\delta_{ij}R+\xi
e^{-i\delta_1} M_1\Gamma^{\ell}_{ij} L]\nu_j + h.c. ,
\vspace{.6cm}
\end{eqnarray}
where $V_L$ is the Cabibbo-Kobayashi-Maskawa matrix; $\Gamma$, $\Gamma'$
are dimensionless $3\times 3$ matrices characterizing the Yukawa
couplings. The small parameter $\xi$ parametrizes the breaking of the
discrete symmetry of the Lagrangian under the $\Phi_1\to \Phi_1,
\Phi_2\to -\Phi_2$ transformations and $\delta_1$  and $\delta_2$ are the
phases that
signal CP-violation in the up- and down-type quark sectors.
$M_{1,2}$,  are mass parameters of the order of the W boson mass, $\tan
\beta\equiv v_2/v_1$ is the ratio of v.e.v.'s for the two Higgs doublets,
and finally $g$ denotes the SU(2) coupling constant.

Since the SM contribution to CP violation and FCNC are very suppressed
for the up-type quark sector, we can make a further simplification [8]
(namely, $\Gamma'=0$) in order to enhance the effects of the charged
Higgs boson to the CP violation and FCNC in the up sector of quarks (in
the following we use $\delta_1=\delta$).
Furthermore, since $\Gamma$ is not {\em a priori} suppressed by the CKM
non-diagonal entries one could expect that $\Gamma$ gives the main
contribution of charged Higgses to processes involving light flavors. As
in Ref. [8] here we will neglect the contributions proportional to $M_u$,
$M_d$, and $M_{\ell}$ in Eq.(1).

The amplitude for $K^+(p_K)\to
\pi^0(p_{\pi})+\mu^+(p_{\mu})+\nu(p_{\nu})$ process contains two pieces
(at tree level):
\vspace{.3cm}
\begin{equation}
\mid {\cal M}\mid^2 =\mid{\cal M}_{SM}+{\cal M}_H\mid^2=\mid{\cal
M}_{SM}\mid^2+\mid{\cal M}_H\mid^2+2 \Re e({\cal M}^*_{SM}{\cal M}_H),
\vspace{.3cm}
\end{equation}
where the standard model contribution is
\vspace{.3cm}
\begin{equation}
{\cal M}_{SM}=\frac {\textstyle G_F}{\sqrt
2}V_{us}\langle\pi^0\mid\overline {s}\gamma_{\mu} u\mid
K^+\rangle\overline {\textstyle
u}(p_{\mu},s_{\mu})\gamma_{\mu}(1+\gamma_5)v(p_{\nu}),
\vspace{.3cm}
\end{equation}
and the charged Higgs boson contribution is given by:
\vspace{.3cm}
\begin{equation}
{\cal M}_H=\frac {\textstyle G_F}{\sqrt
2}\lambda_{us}\lambda_{\mu\nu}e^{2 i \delta}\langle\pi^0\mid{\overline
s}u\mid K^+\rangle\overline {\textstyle u}(p_{\mu},s_{\mu})(1+\gamma_5)v(q),
\vspace{.3cm}
\end{equation}
Here  $\lambda_{ij}=\xi M_1\Gamma_{ij}/m_H$ are dimensionless effective
Yukawa couplings which values were constrained in Ref. [8].

Following Garisto and Kane in Ref. [1] the transverse muon polarization
is given by
\vspace{.3cm}
\begin{eqnarray}
P^{\perp}_{\mu} &= \frac {\textstyle \mid{\cal M}^+\mid^2-\mid{\cal
M}^-\mid^2}{\textstyle \mid{\cal M}^+\mid^2+\mid{\cal
M}^-\mid^2}\nonumber \\[.5cm]
&\simeq \frac {\textstyle 4 {\Re e}({\cal M}^*_{SM}{\cal
M}_H)}{\textstyle \sum_{spins}\mid{\cal M}_{SM}\mid^2}.
\vspace{.3cm}
\end{eqnarray}
The superscripts ($\pm$) refer to the up and down directions of the muon
spin ($s_{\mu}$) respect to the decay plane.

The latter expression in Eq.(5) is obtained from the interference
between the SM and scalar contributions which is proportional to the muon
spin, and using the approximation $\mid {\cal M}_{SM}\mid ^2 \gg \mid
{\cal M}_H\mid ^2$ in the denominator.

The numerator in Eq.(5) is given by
\vspace{.3cm}
\begin{eqnarray}
4\Re e({\cal M}^*_{SM}{\cal M}_H) \simeq {\textstyle 4\sqrt{2} G_F
V_{us}[\displaystyle \frac  {M_K}{m_s}]\frac{f^2_+}{M^2_W}[M_K
\epsilon_{\alpha\beta\gamma\delta}s^{\alpha}p_K^{\beta}p_{\mu}^{\gamma}
p_{\nu}^{\delta}}]{\cal I}m\xi,
\vspace{.3cm}
\end{eqnarray}
where
\begin{equation}
{\cal I}m\xi=\frac{4G_F}{\sqrt2} M^2_W
\lambda_{us}\lambda_{\mu\nu}\sin 2\delta.
\end{equation}
The denominator of Eq.(5) is given by
\vspace{.3cm}
\begin{equation}
\sum_{{\mbox spins}}\mid{\cal M}_{SM}\mid^2=4^2{\displaystyle G^2_F
V_{us}^2 f^2_+} \Phi
\vspace{.3cm}
\end{equation}
where $\Phi$ is the phase space factor
\vspace{.3cm}
\begin{equation}
\Phi= 2(p_{\mu}\cdot p_K)(p_{\nu}\cdot p_K)-M^2_K  p_{\mu}\cdot p_{\nu}+
m^2_{\mu}(-p_{\nu}\cdot p_K+\frac {p_{\mu}\cdot p_{\nu}}{4}),
\vspace{.3cm}
\end{equation}
and $f_+$ is defined from $\langle\pi^0\mid\bar s\gamma_{\mu} u\mid
K^+\rangle\simeq f_+ (p_K+p_{\pi})_{\mu}$.

Putting Eq.(6) and Eq.(8) into Eq.(5) we obtain (see Garisto and Kane in
Ref. [1])
\vspace{.3cm}
\begin{equation}
P^{\perp}_{\mu} \simeq \frac {\sqrt 2}{4}(\displaystyle G_F M^2_W
V_{us})^{-1}[\frac{M_K}{m_s}][\frac{M_K
\epsilon_{\alpha\beta\gamma\delta}{\it
s}^{\alpha}p_K^{\beta}p_{\mu}^{\gamma} p_{\nu}^{\delta}}{\Phi}]{\cal I}m\xi.
\vspace{.3cm}
\end{equation}
As pointed out in Ref. [9], it is convenient to define an average value
for $P^{\perp}_{\mu}$. We define the average value for $P^{\perp}_{\mu}$
in the $K^+$ rest frame as follows:
\vspace{.3cm}
\begin{eqnarray}
\overline {P^{\perp}_{\mu}}=&\hspace{-3.7cm}\int P^{\perp}_{\mu} dp
d\theta\nonumber \\[0.5cm]
&\hspace{-.7cm}\simeq 7.06(\frac {\textstyle 0.199 GeV}{\textstyle
m_s})\lambda_{us}\lambda_{\mu\nu}\sin 2\delta,
\vspace{.3cm}
\end{eqnarray}
where the two independent kinematical variables are taken as
$\displaystyle p\equiv\frac {\mid{\vec p}\mid}{M_K}$ and the angle
$\theta$ ($cos\theta\equiv\hat p_{\mu}\cdot \hat p_{\nu}$).  In the above
numerical result we have neglected terms of order $m^2_{\mu}/M^2_K$ in
the expression for $\Phi$ (as done in Garisto and Kane, Ref. [1]).

As expected, the transverse muon polarization is proportional to
$\delta$, the CP-violating phase that appears in the Yukawa couplings in
Eq.(1). Thus a measurement of $P^{\perp}_{\mu}$ would provide the value
of $\lambda_{us}\lambda_{\mu\nu} \sin 2\delta$.

In order to get a conclusion on the phase $\delta$, we can proceed as
follows. As is pointed in Ref. [10] (p.p. 1530-1531), if we relax the V-A
requirement for the weak charged current responsible for the
$K^+\to\pi^0e^+\nu_e$ process, we can allow in particular an scalar
contribution of the form
\vspace{.3cm}
\begin{equation}
{\cal M}_{S}=\frac {G_F}{\sqrt 2} V_{us} (2M_K) f_S \bar
\ell(1+\gamma_5)\nu_{\ell}.
\end{equation}
If we atribute this amplitude to the exchange of the charged Higgs of our
model, Eq.(12) becomes identical to Eq.(4). This would imply:
\vspace{0.3cm}
\begin{equation}
\mid \lambda_{us}\lambda_{e\nu}\mid\simeq 2M_K V_{us} \big(\frac
{m_s-m_u}{M^2_K-m^2_{\pi}}\big) \big |\frac {f_S}{f_+}\big |.
\vspace{.3cm}
\end{equation}

If we take the experimental value reported in Ref. [11], namely \\
$\mid \frac {\textstyle f_S}{\textstyle f_+}\mid=0.084\pm 0.023$ and if
we asume $e-\mu$ universality, we would obtain:
\vspace{0.3cm}
\begin{equation}
\mid \lambda_{us}\lambda_{\mu\nu_{\mu}}\mid\simeq (1.6\times
10^{-2})\big(\frac {m_s}{0.199 GeV}\big),
\vspace{.3cm}
\end{equation}
which is consistent with the upper bound obtained in Ref. [8].
Using Eq.(14) in Eq.(11), we get
\vspace{.3cm}
\begin{equation}
\overline P^{\perp}_{\mu}\simeq (0.11) \sin 2\delta.
\vspace{.3cm}
\end{equation}
Thus, a measurement of $\overline P^{\perp}_{\mu}$ of the order of
$10^{-3}-10^{-4}$ would imply an small phase, $\delta\sim {\cal
O}(10^{-2}-10^{-3})$.
In fact, since $\sin 2\delta\leq 1$, we expect $\overline
P^{\perp}_{\mu}\leq 10^{-1}$.\\[1em]

Summarizing, we have calculated the transverse muon polarization in the
framework of a general two-Higgs doublet model with spontaneous violation
of CP. Using present bounds on the absolute value of the effective Yukawa
couplings of the charged Higgs boson (Ref.[8] and Eq.(14)), we estimate
$\overline P^{\perp}_{\mu}\simeq 0.11 \sin 2\delta$. Thus, a measurement
of $\overline P^{\perp}_{\mu}$ at the level of $10^{-3}-10^{-4}$, would
imply $\delta\sim {\cal O}(10^{-2}-10^{-3})$. We would like to emphasize
that  $\overline P^{\perp}_{\mu}$ can not be induced from an scalar
amplitude arising from two-Higgs doublet model where we impose the
symmetry under $\Phi_1\to\Phi_1$ and $\Phi_2\to -\Phi_2$ (namely $\xi=0$
in Eq.(1)).\\[1em]
{\bf Acknowledgements}\\[1em]
I would like to thank G. L\'opez Castro for suggesting this problem and
for useful discussions.
\

\end{document}